\documentclass[journal]{IEEEtran}
\IEEEoverridecommandlockouts

\usepackage{xfrac}
\usepackage{amsmath}
\usepackage{multirow,etoolbox}
\usepackage{color}
\usepackage{amssymb}
\usepackage{cite}

\usepackage{diagbox}

\usepackage{subcaption}
\usepackage[utf8]{inputenc}

\newcommand{\mathbbm}[1]{\text{\usefont{U}{bbm}{m}{n}#1}} 
\usepackage{stfloats}
\usepackage{relsize}
\newcommand{\Sum}{\mathlarger{\sum}}

\makeatletter
\patchcmd{\@maketitle}
  {\addvspace{0.5\baselineskip}\egroup}
  {\addvspace{-.9\baselineskip}\egroup}
  {}
  {}
\makeatother

\begin{document}
\bstctlcite{IEEEexample:BSTcontrol}

\title{\makebox[\linewidth]{\parbox{\dimexpr\textwidth+1.5cm\relax}{\centering Multi-user Joint Maximum-Likelihood Detection in Uplink NOMA-IoT Networks: Removing the Error Floor}}}
\author{Hichem Semira, Ferdi Kara,~\IEEEmembership{Senior Member,~IEEE,} Hakan Kaya, Halim Yanikomeroglu,~\IEEEmembership{Fellow,~IEEE} 
\thanks{The work of F. Kara and H. Kaya is supported by Zonguldak Bulent Ecevit University with the project number 2021-75737790-02. H. Semira is with Laboratory of Electronics and New Technologies (LENT), University of Oum El Bouaghi, Oum El Bouaghi, Algeria, email: hichem.semira@univ-oeb.dz. F. Kara and H. Kaya are with the Electrical-Electronics Engineering, Zonguldak Bulent Ecevit University, Zonguldak, Turkey,  e-mail: \{f.kara, hakan.kaya\}@beun.edu.tr. F. Kara and H. Yanikomeroglu are with the Department of Systems and Computer Engineering, Carleton University, Ottawa, K1S 5B6, ON, Canada, e-mail: halim@sce.carleton.ca.}
        }
\maketitle
\begin{abstract}
The Internet of Things (IoT) framework requires a massive number of connection thus demanding spectral efficient solutions such as Non-Orthogonal Multiple Access (NOMA). However, the main drawback of NOMA with successive interference canceler (SIC)-based detectors is the error floor in the uplink. In this paper, a reliable multi-user detection in uplink IoT NOMA is guaranteed by a Joint Maximum-Likelihood (JML) detector (i.e., optimum detection algorithm). We derive a closed-form upper bound of bit error rate (BER) of JML over Rayleigh fading channels for arbitrary number of IoT devices and an adaptive M-ary phase shift keying (M-PSK). Based on the extensive simulations, the derived expressions are validated and it is revealed that the JML improves the error performance in uplink NOMA and removes the error floor. Furthermore, regardless of the number of the IoT devices and modulation order, a full diversity order (i.e., number of receiving antennas) is guaranteed for each device.

\end{abstract}

\IEEEpeerreviewmaketitle
\section{Introduction}
Non-Orthogonal Multiple Access (NOMA) scheme allows several users sharing entirely the same available time/frequency resources. Accordingly, with the aim to support massive machine-type communication (MMTC) in Internet of Things (IoT) framework, NOMA has emerged as an appropriate multiple access scheme \cite{Mostafa2019,Emir2021}. In NOMA, the successive interference cancellation (SIC) technique is consistently considered to eliminate inter-user interference (IUI). The performances of NOMA systems for SIC detector are studied in-depth in terms of information-theoretic perspectives (e.g., outage probability and capacity \cite{Tegos2020,Ding2020}), which are based on theoretical Shannon limit. However, most of the existing works consider perfect SIC \cite{Ding2020} which is not reasonable for practical scenarios. In \cite{Tegos2020}, the authors relax this assumption and propose a joint decoding to remove the error floor of outage probability in the uplink NOMA due to SIC detectors. Nevertheless, this SINR-based analysis can reveal the performance for finite alphabets (i.e., short-packet communication) and it is proved that the SIC is not optimal in uplink NOMA for infinite alphabets \cite{Kara2018d,Kara2020}. It noticeably suffers from the error floor and has a severe performance (e.g., bit error rate (BER)). In fact, this drawback gets worsen when the number of users is increased (e.g., more than two users), and none of the symbols becomes detectable.
To resolve the error performance problem with infinite alphabets in the uplink, the joint maximum likelihood-detector (JML) has recently attracted attention \cite{Yeom2019,Kara2020b,Shahab2020,Semira2021}. These studies have shown the predominance of JML over SIC in the uplink. Besides, the JML is still the optimal detection for non-fully overlapped waveforms (e.g., semi-orthogonal) \cite{Shahab2021}. However, these studies are mostly based on simulations whereas an upper bound analysis of JML detector is investigated only in \cite{Yeom2019,Semira2021}. Nevertheless, these studies are devoted to special cases by considering only two users which again do not meet the requirements of IoT networks. Thus, to enable the MMTC with a reliable performance, the JML in the multi-user uplink NOMA is still to be explored.

Motivated by the above discussions, in this letter, we study an uplink NOMA for an IoT network with a JML detector to guarantee a reliable multi-user detection. To the best of our knowledge, this work is the first to derive an upper bound BER expression for an arbitrary number of IoT devices and with an adaptive M-PSK modulation (according to link quality \cite{3gpp1630}). The analytical expressions are validated via computer simulations and we reveal that the JML eliminates the error floor completely and a full diversity order (i.e., receiving antenna) is achieved regardless of number of IoT devices or modulation order. 

The rest of the paper is organized as follows. In Section II, the system model is introduced. In Section III, we perform an error performance analysis and an upper bound of BER is derived. Later, in Section IV, we validate the analytical results via simulations. Finally, Section V concludes the paper.
\section{System Model}
We consider an uplink NOMA scheme, where $K$ active IoT devices belonging to the same cluster are sending their data towards a single base station (BS). We assume that the BS is equipped with $L$ antennas and the IoT devices are equipped with a single antenna (SIMO-NOMA). The $K$ IoT devices are allowed with equal priority to simultaneously access the same frequency block using their own powers. Hence, the received signal at the BS is given by
\begin{equation}\label{eq:1}
\mathbf{y}=\sum_{{k}=1}^{K} \mathbf{c}_k \sqrt{P_{k}} x_k +\mathbf{w},   \end{equation} 
where $\mathbf{y}\in\mathbb{C}^{L\times1}$, $P_{k}$ is the transmitted power for the $k^{th}$ device (D$_k$) and $\mathbf{c}_k\in\mathbb{C}^{L\times1}$  denotes the flat fading channel vector (counting effects of path loss and small-scale fading) between the $k^{th}$ device and the BS. The components of the vector $\mathbf{c}_k$ are independent and identically distributed (i.i.d.) and follow $\mathbf{c}_k \sim \mathcal{CN}(0,\sigma_k^2\mathbf{I}_L)$, where  $\mathbf{I}_L$ denotes the $L \times L$ identity matrix. The additive white Gaussian noise (AWGN) vector $\mathbf{w}\in\mathbb{C}^{L\times1}$  is given  by \small$\mathbf{w} \sim \mathcal{CN}(0,\frac{N_0}{2}\mathbf{I}_L)$. \normalsize At the BS a channel state information (CSI) for each device is supposed to be available, and the received signal powers are presumed to be correctly measured using the pilot signals, such that
 $ \left\|\mathbf{h}_1\right\|> \left\|\mathbf{h}_2\right\|>  \cdots >\left\|\mathbf{h}_K\right\|$, where $\mathbf{h}_k=\sqrt{P_{k}}\mathbf{c}_k$, $k=1,\cdots\,K$, and the symbol $\left\|.\right\|$ denotes the Frobenius norm.
 
The information bitstream is modulated as complex symbol $x_k\in \mathbb{C}$, $k=1, \cdots,K$, where it takes its value from the alphabet $\mathbf{\chi}_{k}=\left[ s_{k1},s_{k2},\cdots,s_{kM_k}\right]^T$, and the symbol $s_{kn_k}$ is the $n_k^{th}$ constellation point in $M_k$-ary modulation order of D$_k$. $|s_{kn_k}|= \sqrt{\epsilon_k}=\sqrt{\epsilon_{b}{\log_{2}}{M_{k}}}$  for D$_k$, where $\epsilon_k$ is the symbol energy and $\epsilon_{b}$ is the bit energy.
We consider an adaptive modulation that uses $M_k$-PSK with Gray coded mapping.
\subsubsection{MRC-SIC detector}
In SIC detector, IoT devices' symbols are detected serially. Initially, the symbols of the IoT device with strongest gain channel is detected. Then, the contribution of these detected symbols to received signal is removed, producing a second signal to detect the IoT device's symbols with second-best channel condition. This process is repeated until the symbols of the last IoT device with weakest channel condition are detected. Consequently, the procedure is summarized as
\begin{equation} \label{eq:2}
\Hat{x}_k = \underset{s_{kn_k}}{\mathrm{argmin}}\left\|\mathbf{y}_{SIC,k}-\mathbf{h}_{k} s_{kn_k}\right\|^2,
\end{equation} 
where $k=2, \cdots, K$, $n_{k}=1,\cdots, M_k$, and
\begin{equation} \label{eq:2bis}
\mathbf{y}_{SIC,k} = \mathbf{y}_{SIC,k-1}-\mathbf{h}_{k-1}\Hat{x}_{k-1},
\end{equation} 
where $\mathbf{y}_{SIC,1}=\mathbf{y}$, and $\Hat{x}_1 = \underset{s_{1n_1}}{\mathrm{argmin}}\left\|\mathbf{y}-\mathbf{h}_{1} s_{1n_1}\right\|^2$.

Although the procedure seems simple to implement, unfortunately, it suffers from the error floor. To recover each signal, the detector treats the remaining low power signals as interference, thus an error floor occurs. Increasing the number of antennas at the receiver may improve the performance of SIC in low SNR, but can not remove the error floor \cite{Kara2020}. 
\subsubsection{MRC-JML detector}
To avoid the drawback of the SIC detector, we suggest to jointly recover (i.e., optimal detection) the transmitted signals from IoT devices. The JML detector accomplishes an exhaustive search to simultaneously decode all signals $x_1, x_2,\cdots,x_k$ as follows 
\begin{equation} \label{eq:3}
\lbrack\Hat{x}_1,\Hat{x}_2,\cdots,\Hat{x}_k\rbrack= \underset{s_{kn_k}}{\mathrm{argmin}}\Big \|\mathbf{y}-\sum\nolimits_{{k}=1}^{K} \mathbf{h}_k s_{kn_k}\Big \|^2.
\end{equation} 

Regardless of the order of selection, the JML traits the signals the same  by realizing a full joint search for all possible combinations $\lbrack \mathbf{h}_1 s_{1n_{1}},\mathbf{h}_2 s_{2n_{2}},\cdots,\mathbf{h}_K s_{Kn_{K}} \rbrack$, which avoids an error floor. 
\subsubsection{Computational complexity}
The computational complexity comparison for MRC-JML and MRC-SIC is given in Table \ref{Tab:tb1}. The complexity of MRC-JML increases exponentially with the increase of $K$ and $M_k$ whereas the complexity of MRC-SIC increases linearly with the same factors. However, this complexity reflects the required computational capacity rather than computing time. One can easily see from (4) that the search in JML is independent so that it can be computed in parallel and be compared in the final stage, whereas the SIC detection (2)-(3) should be implemented sequentially. Therefore, as long as we have enough computational capacity, which is quite possible since it is an uplink communication and the JML is implemented at the BS, the JML does not cause a latency due to computational time. 
Besides, this computational burden can be considered  as acceptable, especially if we take into account the performance improvement by MRC-JML.
\begin{table} 
\centering
\caption{Computational cost in terms of real operations}
\addtolength{\tabcolsep}{-5.5pt}
\scalebox{0.95}{%
   \begin{tabular} {c|c|c|c|}      
   \hline
     &  Multiplier & Adder & Comparator \\
      \hline 
     SIC  & $6L\sum\limits_{{i}=1}^{K}M_{i}+4L(K-1)$  & $(6L-1)\sum\limits_{{i}=1}^{K}M_{i}+4L(K-1)$  & $\sum\limits_{{i}=1}^{K}(M_{i}-1)$      \\
    \hline
     JML  & $(4LK+2L)\prod\limits_{i=1}^{K}M_{i}$  & $(4LK+2L-1)\prod\limits_{i=1}^{K}M_{i}$ & $\prod\limits_{i=1}^{K}M_{i}-1$      \\
\hline
\end{tabular}}
\label{Tab:tb1}
\end{table}

\section{Error Performance Analysis}
In this section, we analyse the error performance of IoT uplink-NOMA system with JML. We derive an analytic expression of the upper bound for the BER in fading channels. For the purpose of a comprehensive study, we assume that the IoT device of order $k$ sends its data according to an appropriate order modulation $M_k$, which depends on the channel quality. 

For the representation simplicity, let us start by considering a system composed of three IoT devices. The Fig. 1 shows the superposed symbols, where the D$_1$ chooses its symbols from 8-PSK constellation $(M_1=8)$ and both the others devices: D$_2$ and D$_3$ from a quadrature phase constellation  $(M_2=4$ and $M_3=4)$. It is worth mentioning that these constellation orders depend to the channel quality. The first three bits\footnote{For an easy-follow illustration, we have adopted an octal representation of the symbols for D$_1$, and quaternary representation for D$_2$ and D$_3$.} belong to D$_1$, the following two bits belong to D$_2$ and the last two bits belong to D$_3$. Thus, the binary bit representations of the three symbols are given in the form of ${\{b_{11} b_{12} b_{13} b_{21} b_{22} b_{31} b_{32}\}}$, where the first sub-index represents the device and the second represents the bit order within the symbol.

\begin{figure*}
\centering
\subfloat[{Error pattern for Device1 (D$_1$)}]{\includegraphics[width=0.66\columnwidth]{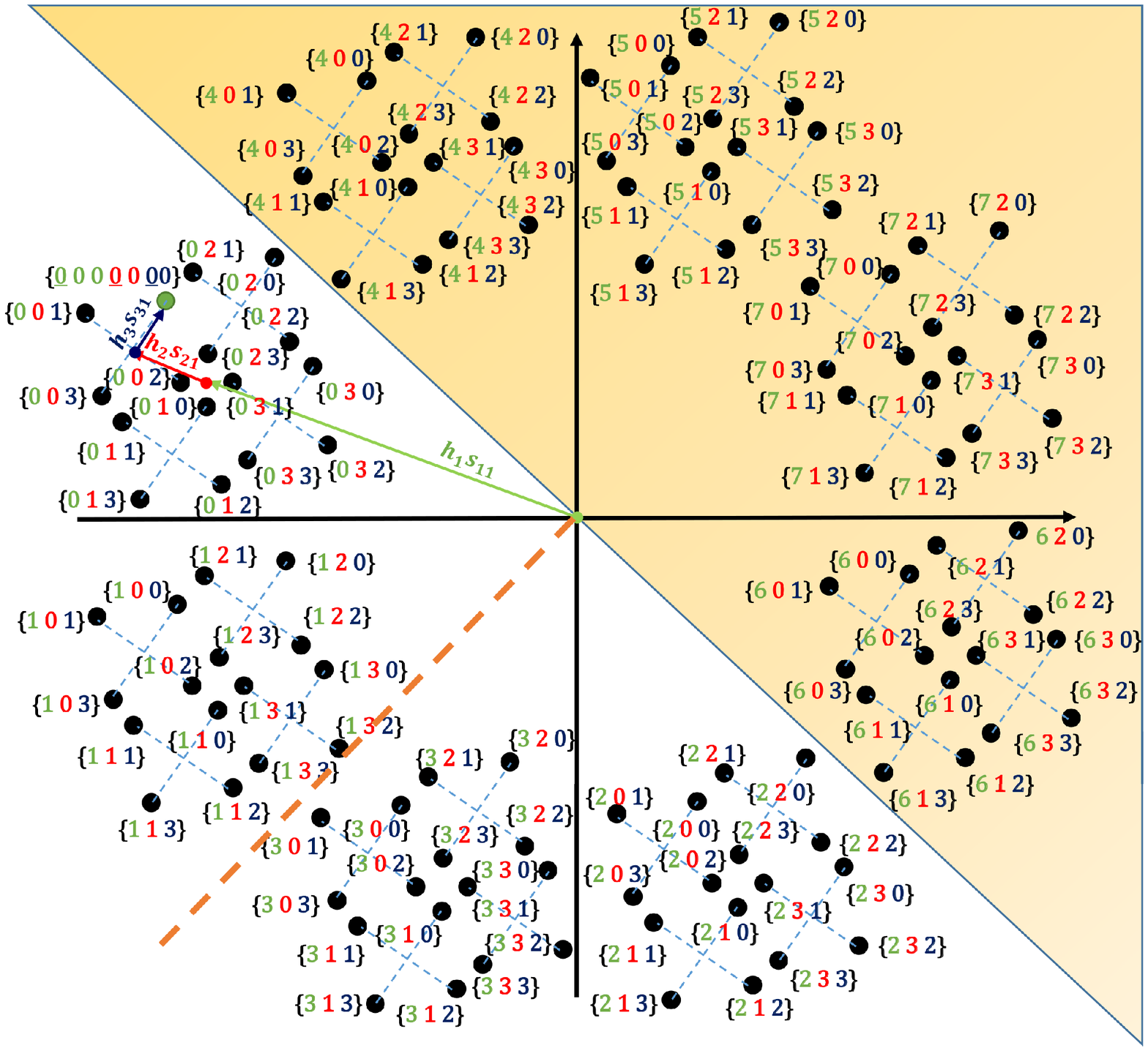}
\label{img1:const1}}
\subfloat[{Error pattern for Device2 (D$_2$)}]{\includegraphics[width=0.66\columnwidth]{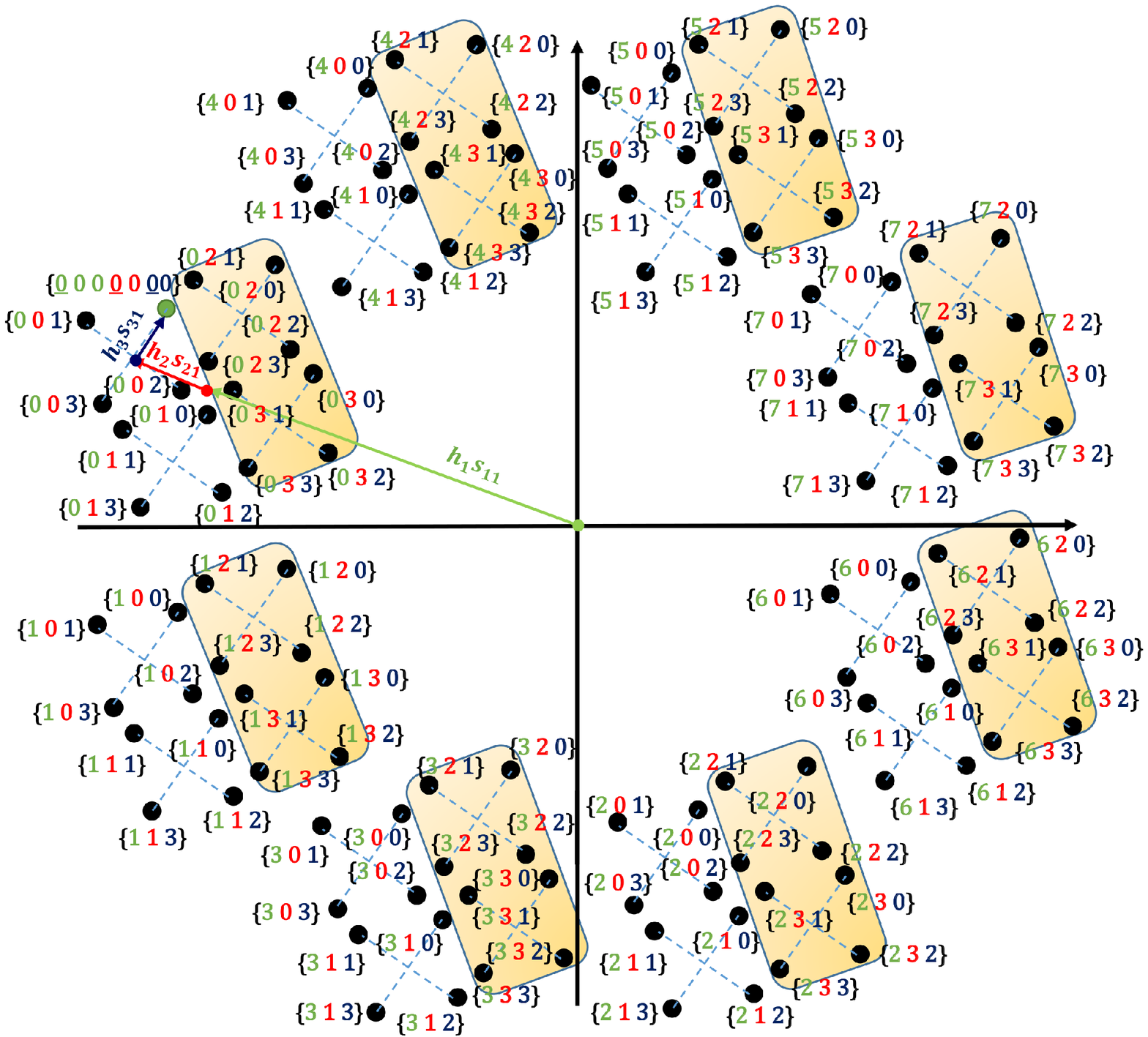}
\label{img2:const2}}
\subfloat[{Error pattern for Device3 (D$_3$)}]{\includegraphics[width=0.66\columnwidth]{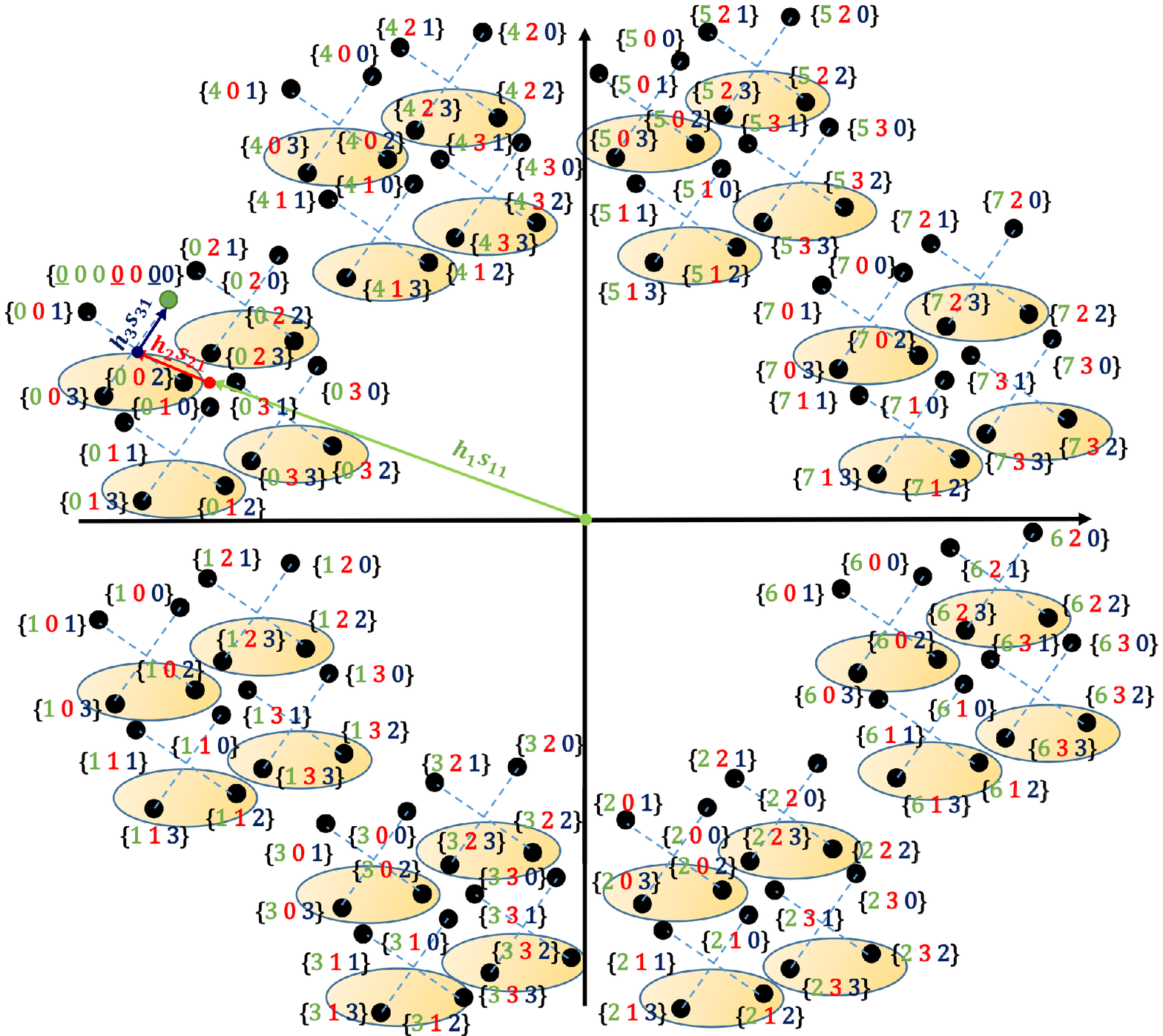}
\label{img3:const3}}
\caption{{Signal space diagram for superposed symbols of three IoT devices (8-PSK with two 4-PSK) with Gray coded bit mapping.}}
\label{constellations}
\end{figure*}
\begin{figure}
    \centering
    \includegraphics[width=0.65\columnwidth]{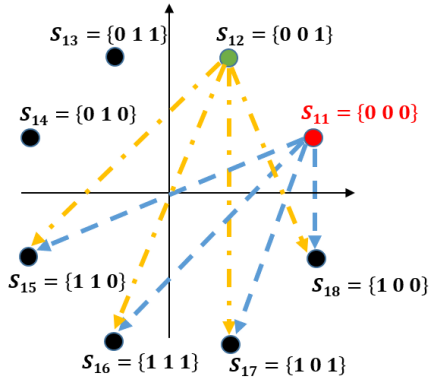}
    \caption{The effective distances for D$_1$.}
    \label{fig2:fig4}
\end{figure}

For convenience, we assume that the superimposed sequence $\{000, 00, 00\}$ that matches the symbols $x_1=$ $s_{11}=\exp\left(j \frac{\pi}{8} \right)$ for D$_1$, $x_2=$ $s_{21}= \exp\left(j \frac{\pi}{4} \right)$ for D$_2$ and $x_3=$ $s_{31}= \exp\left(j \frac{\pi}{4} \right)$ for D$_3$ is received at the BS. Then, we proceed to analyze the BER of the first bit ($b_{k1}=0$) for each of the three devices. In view of symmetry of M-PSK modulation, it is evident that the BER is unchanged when $b_{k1}=1$, therefore, the study is restricted on the case of $b_{k1}=0$. Since the BS relies on ML detector to detect the symbols, it is clear that an error happens if the ML selects the device symbol with first bit $b_{k1}=1$. Fig. \ref{img1:const1} shows the location of the erroneous symbols for D$_1$ (shaded area) when the JML selects one of the superimposed symbols with the first bit equal to one, i.e., the lower half-circle of an 8-PSK constellation with Gray coded bit mapping (see Fig. \ref{fig2:fig4}). It is simple to notice that the number of erroneous symbols after this wrong decision is equal to the all possible symbols of the devices D$_2$ and D$_3$ superimposed with the four symbols of D$_1$ ($\frac{8}{2}\times4\times4$ symbols). Thus, if we consider a K IoT devices with $M_k$-ary order for each D$_k$, the number of erroneous symbols is clearly equal to $\frac{1}{2}\prod_{i=1}^{K} M_{i}$. Regarding the others devices (the remaining $K-1$ devices), the same number of errors can be noticed, see Fig. \ref{img2:const2} and Fig. \ref{img3:const3} for illustration. By examining the error patterns of the Fig. \ref{img2:const2} and Fig. \ref{img3:const3}, we observe the same number of errors which equal to $8\times \frac{4}{2} \times4$ and $8 \times 4 \times \frac{4}{2}$  for D$_2$ and D$_3$, respectively. The difference between errors of distinct devices is due to error pattern that depends to the arrangement of the components of signals (power) within the overlapped symbols at receiver.

Based on the result above, we designate the distances between the first symbol $s_{k1}$ of each IoT D$_k$, $k=1,\cdots,K$, and the $\frac{M_k}{2}$ erroneous symbols $s_{kn_{k}}$, $n_{k}= \frac{M_k}{2}+1,\cdots, M_k$, by the vector $\mathbf{E}_{k1}=\sqrt{\epsilon_k}^{-1}\left[s_{k1}-s_{k{\frac{M_k}{2}+1}},\cdots,s_{k1}-s_{kM_{k}}\right]^T$, and the distances between $s_{k1}$ with the symbols of the alphabet $\mathbf{\chi}_{k}$ by the vector $\mathfrak{d}_{k}=\sqrt{\epsilon_k}^{-1}(s_{k1}\otimes\mathbbm{1}_{M_k} -\mathbf{\chi}_{k})$, where $\otimes$ denotes the Kronecker product, and  $\mathbbm{1}_{M_k}$ is the all ones vector of dimension $M_k\times1$. Correspondingly, the total $\frac{1}{2}\prod_{i=1}^{K} M_{i}$ pairwise error probabilities (PEP) following the erroneous decision between the symbol $s_{k1}$ with the symbols $s_{kn_{k}}$, $n_{k}=\frac{M_k}{2}+1,\cdots,M_k$, conditioned on the vectors $\textbf{h}_k$, $k=1,\cdots,K$ are given by
\small
\begin{equation} \label{Eq4} 
  Pr\left({s_{k1}\to s_{kn_{k}}} \mid {\mathbf{h}_1,\cdots,\mathbf{h}_K}\right) = Q\left(\frac{\parallel \sum_{i=1}^{K}  {\mathbf{h}}_{i}\sqrt{\epsilon_i}{\mathbf{d}}_{i}^{D_k}(m)\parallel
}{\sqrt{2  N_0}}\right),
\end{equation}
\normalsize
where \small $m=1,\cdots,\frac{1}{2}\prod_{i=1}^{K} M_{i}$, \normalsize and \small $\mathbf{d}_{i}^{D_k}$ $\in\mathbb{C}^{\left(\frac{1}{2}\prod_{i=1}^{K} M_{i}\right)\times1}$, $i=1,\cdots,K$ \normalsize, denotes the distances between the symbol under test $s_{k1}$ and the superimposed symbols with the erroneous symbol (first bit $b_{k1}=1$). Table \ref{Tab:tb2} summarizes the elements of $\mathbf{d}_{i}^{D_k}$. It is noteworthy that the distances $\parallel \sum_{i=1}^{K}  {\mathbf{h}}_{i}\sqrt{\epsilon_i}{\mathbf{d}}_{i}^{D_k}\left(m\right)\parallel$ are the distances between $M_k$-PSK symbols in a new space rotated by $\mathbf{h}_k$, $k=1,\cdots,K$, and as long as the symbols have experienced the same rotation, in consequence, the symmetry and error patterns are retained. As a result of these properties and the circularity of PSK modulation, if we replace the term $s_{k1}$ by any other terms $s_{kn_{k}}$, where $n_{k}=2,\cdots,M_k$, the elements of $\mathfrak{d}_{k}$ are the same without dealing with order, then the components of $\mathbf{d}_{i}^{D_k}$ for $i\neq k$, which contribute to the conditional PEP turns into the same. This means that the conditional PEP for different IoT devices, differs from each other in virtue of vectors $\mathbf{E}_{k1}$.
\begin{table*} 
\centering
\caption{Distances to erroneous symbols}
\addtolength{\tabcolsep}{-4pt}
\scalebox{1}{%
   \begin{tabular} {|c||c|c|c|c|c|c|}       
   \hline
   \backslashbox[25 mm]{Devices}{Distances}
  & 
  $\mathbf{d}_{1}^{D_{k}}$ & $\mathbf{d}_{2}^{D_{k}}$ & $\cdots$ & $\mathbf{d}_{k}^{D_{k}}$ & $\cdots$& $\mathbf{d}_{K}^{D_{k}}$\\
      \hline \hline  
     D$_1$  & $\mathbf{E}_{11} \otimes \mathbbm{1}_{\prod_{i=2}^{K}M_{i}}$  &
    $\mathbbm{1}_{\frac{M_{1}}{2}}\otimes \mathfrak{d}_{2}\otimes \mathbbm{1}_{ \prod_{i=3}^{K}M_{i}}$   & 
     $\cdots$  & 
     $\mathbbm{1}_{\frac{1}{2}\prod_{i=1}^{k-1} M_{i}} \otimes \mathfrak{d}_{k}\otimes \mathbbm{1}_{ \prod_{i=k+1}^{K}M_{i}}$  &
     $\cdots$  &
     $\mathbbm{1}_{\frac{1}{2}\prod_{i=1}^{K-1} M_{i}} \otimes \mathfrak{d}_{K}$  \\
   \hline
     D$_2$  & $ \mathfrak{d}_{1}\otimes \mathbbm{1}_{\frac{1}{2} \prod_{i=2}^{K}M_{i}}$  & 
     $\mathbbm{1}_{M_{1}}\otimes \mathbf{E}_{21}\otimes \mathbbm{1}_{\prod_{i=3}^{K}M_{i}}$  & $\cdots$  & 
     $\mathbbm{1}_{\frac{1}{2}\prod_{i=1}^{k-1} M_{i}} \otimes \mathfrak{d}_{k}\otimes \mathbbm{1}_{ \prod_{i=k+1}^{K}M_{i}}$ & $\cdots$ & 
     $\mathbbm{1}_{\frac{1}{2}\prod_{i=1}^{K-1} M_{i}} \otimes \mathfrak{d}_{K}$  \\
     \hline \hline  
     D$_k$  & $ \mathfrak{d}_{1}\otimes \mathbbm{1}_{\frac{1}{2} \prod_{i=2}^{K}M_{i}}$  &
     $\mathbbm{1}_{M_{1}}\otimes \mathfrak{d}_{2}\otimes \mathbbm{1}_{\frac{1}{2} \prod_{i=3}^{K}M_{i}}$  & $\cdots$ &
     $\mathbbm{1}_{\prod_{i=1}^{k-1} M_{i}} \otimes \mathbf{E}_{k1}\otimes \mathbbm{1}_{ \prod_{i=k+1}^{K}M_{i}}$ & $\cdots$ & 
     $\mathbbm{1}_{\frac{1}{2}\prod_{i=1}^{K-1} M_{i}} \otimes \mathfrak{d}_{K}$  \\
     \hline \hline  
    D$_K$  &  $\mathfrak{d}_{1}\otimes \mathbbm{1}_{\frac{1}{2} \prod_{i=2}^{K}M_{i}}$ & 
   $\mathbbm{1}_{M_{1}}\otimes \mathfrak{d}_{2}\otimes \mathbbm{1}_{\frac{1}{2} \prod_{i=3}^{K}M_{i}}$ & $\cdots$ &
   $\mathbbm{1}_{\prod_{i=1}^{k-1} M_{i}} \otimes \mathfrak{d}_{k}\otimes \mathbbm{1}_{\frac{1}{2} \prod_{i=k+1}^{K}M_{i}}$ &
   $\cdots$ &
  $\mathbbm{1}_{\prod_{i=1}^{K-1} M_{i}} \otimes \mathbf{E}_{K1}$   \\
\hline
\end{tabular}}
\label{Tab:tb2}
\end{table*}

At this juncture, we have study the error patterns for symbol $s_{k1}$, it is easy to show that the number of errors is constantly the same for the remaining of symbols $s_{kn_{k}}$, $n_{k}=2,\cdots,M_{k}$. From the symmetry of the M$_k$-PSK constellation (see Fig. \ref{fig2:fig4}), we can notice  that the  distances between the points which are located in upper half-circle and those situated in lower half-circle are equal in pairs with respect to the right and left semicircle, consequently, there are a total of $\frac{M_k}{4}$ different vectors $\mathbf{d}_{k}^{D_{k}}$ which depend to $\mathbf{E}_{kn}=\sqrt{\epsilon_k}^{-1}\left[s_{kn}-s_{k{\frac{M_k}{2}+1}},\cdots,s_{kn}-s_{kM_{k}} \right]^T$, \small $n=1,\cdots,\frac{M_k}{4}$ \normalsize. Thus, we consider only the contribution of $\frac{M_k}{4}$ symbols located in the right quadrant to calculate the upper bound. Additionally, by examining the remaining of bits $b_{k{\log_{2}}M_{k}}$ from symbols $s_{kn}$, \small $n=1,\cdots,\frac{M_k}{4}$ \normalsize, and taking into consideration the approximated formula of the probability of error \cite[Eq. (10)]{Jianhua99} which ties to the right quadrant, i.e., utilizing only the $\frac{M_k}{4}$ symbols\footnote{The analysis for $M_k = 2$ is identical to $M_k = 4$.}, and by supposing equally likely symbols in the new space diagram, we get
\small
\begin{equation} \label{eq:5} 
Pr\left(e\mid {\mathbf{h}_1,\cdots,\mathbf{h}_K}\right) \cong \frac{M_k}{2\log_2{M_k}}P_{rq},
\end{equation}
\normalsize
where
\small
\begin{equation} \label{eq:6}
P_{rq}=\frac{4}{M_k}\Sum_{{n}=1}^{\frac{M_k}{4}}Q\left(\frac{\parallel {\mathbf{h}}_{k}\sqrt{\epsilon_k}{\mathbf{d}}_{kn}^{D_k}(m)+\sum_{\substack{i=1 \\ i\neq k}}^{K}  {\mathbf{h}}_{i}\sqrt{\epsilon_i}{\mathbf{d}}_{i}^{D_k}(m)\parallel
}{\sqrt{2  N_0}}\right),
 \end{equation} 
 \normalsize
where  \small$\mathbf{d}_{kn}^{D_k}=\mathbbm{1}_{\prod_{i=1}^{k-1} M_{i}} \otimes \mathbf{E}_{kn}\otimes\mathbbm{1}_{\prod_{i=k+1}^{K}M_{i}}$\normalsize. Incorporating (\ref{eq:6}) into (\ref{eq:5}), yields 
\begin{equation}\label{eq:7} 
 Pr\left(e\mid {\mathbf{h}_1,\cdots,\mathbf{h}_K}\right) \cong\\ \frac{2}{\log_2{M_k}}\sum_{{n}=1}^{\frac{M_k}{4}}Q\left(\frac{\parallel \delta_{nm}^{D_k} \parallel}{\sqrt{2  N_0}}\right),
 \end{equation}
where \small $\delta_{nm}^{D_k}={\mathbf{h}}_{k}\sqrt{\epsilon_k}{\mathbf{d}}_{kn}^{D_k}(m)+\sum_{\substack{i=1 \\ i\neq k}}^{K} {\mathbf{h}}_{i}\sqrt{\epsilon_i}{\mathbf{d}}_{i}^{D_k}(m)$\normalsize. Hence, for the probability of union of events, we write the upper bound by
  \small
\begin{equation}\label{eq:8} 
 Pr\left(e\mid {\mathbf{h}_1,\cdots,\mathbf{h}_K}\right) \leq\frac{2}{\log_2{M_k}}\sum_{{n}=1}^{\frac{M_k}{4}} \sum_{{m}=1}^{\frac{1}{2}\prod_{i=1}^{K} M_{i}}     Q\left(\frac{\parallel \delta_{nm}^{D_k}\parallel
}{\sqrt{2  N_0}}\right).
\end{equation}
\normalsize
Due to the characteristic of the vectors $\mathbf{h}_1,\cdots,\mathbf{h}_k$, we have \small$\delta_{nm}^{D_k}\sim   \mathcal{CN} \left( 0,P_{k}\epsilon_k\sigma_{k}^{2} |{\mathbf{d}_{kn}^{D_k}}(m)|^2+\sum_{\substack{i=1 \\ i\neq k}}^{K} P_{i}\epsilon_i\sigma_{i}^{2}|\mathbf{d}_{i}^{D_k}(m)|^2 \right)$\normalsize. Like \cite{Yeom2019}, if we denote by \small $\Gamma_{km}=\frac{{\parallel \delta_{nm}^{D_k}\parallel
}^2}{2 N_0}$ \normalsize and considering $L$ branches of uncorrelated signals $\textbf{y}\in\mathbb{C}^{L\times1}$ received at BS, then $\Gamma_{km}$ follows Erlang distribution with parameter \small $\Upsilon_{nm}^{D_k}=\frac{1}{2}\left( \zeta_{k} |{\mathbf{d}_{kn}^{D_k}}(m)|^2+\sum_{\substack{i=1 \\ i\neq k}}^{K}  \zeta_{i}|\mathbf{d}_{i}^{D_k}(m)|^2 \right)$ \normalsize and PDF given as
\begin{equation} \label{eq:9}
 P_{\Gamma_{km}}\left(\gamma\right)=\frac{1}{\left(L-1\right)!}\frac{\gamma^{L-1}}{{\left(\Upsilon_{nm}^{D_k}\right)}^L} \exp{\left(\frac{-\gamma}{\Upsilon_{nm}^{D_k}}\right)}.
 \end{equation}
where \small$\zeta_{k}=\frac{P_k\epsilon_k\sigma_k^2}{N_0}$ \normalsize is the received signal-to-noise ratio (SNR$_k$) of D$_k$. Therefore, by averaging the BER $P_{k}{\left(e\right)}$ of IoT D$_k$ over the distribution of $\Gamma_{km}$, it becomes
\small
\begin{equation} \label{eq:10}
 P_{k}{\left(e\right)}\leq \frac{2}{\log_2{M_k}}\Sum_{{n}=1}^{\frac{M_k}{4}} \Sum_{m=1}^{\frac{1}{2}\prod_{i=1}^{K} M_{i}}\int_{0}^\infty Q\left(\sqrt{\gamma}\right)
 P_{\Gamma_{km}}\left(\gamma\right) d \gamma.  
\end{equation}
\normalsize
With the aid of \cite{Yeom2019}, we obtain (\ref{eq:11}) as the upper bound \small $U_{P_k(e)}$\normalsize, where \small $P_{k}{\left(e\right)}\leq U_{P_k(e)}$\normalsize. The expression (\ref{eq:11}) is the general form of BER upper bound for the $k^{th}$ IoT device where $k=1,\cdots,K$, using $M_k$-ary PSK. From Fig. \ref{img1:const1} and Fig. \ref{fig2:fig4}, we can deduce that the vector $\mathbf{E}_{k1}$ contains the smallest distances (with the minimum norm) in comparison with those obtained in the same quadrant; hence, their conditional PEP \small $Pr\left({{s_{k1}}\to {s_{kn_{k}}}\mid {\mathbf{h}_1,\cdots,\mathbf{h}_K}}\right)$ \normalsize dominate the BER for high SNR. Then,  for $M_k>4$ the upper bound (\ref{eq:11}) may be approximated using only $\mathbf{E}_{k1}$ to get (\ref{eq:12}). It can be noticed that equations (\ref{eq:11}) and (\ref{eq:12}) are monotonically decreasing functions in terms of superimposed SNR$_k$ of different IoT devices, which determines the performance of each device according to its quality of the channel and the transmission power with the order of modulation.
\begin{figure*}
    \begin{equation} \label{eq:11}
U_{P_k(e)}=\frac{1}{\max\left(2,\log_2{M_k}\right)} \sum_{{n}=1}^{\max\left(1,\frac{M_k}{4}\right)} \sum_{m=1}^{\frac{1}{2}\prod_{i=1}^{K} M_{i}} \left[ 1-\sum\nolimits_{l=0}^{L-1} {\binom{2l}{l}} {\left({1+2/\Upsilon_{nm}^{D_k}}\right)^{-\frac{1}{2}}} {\left(2\Upsilon_{nm}^{D_k}+4\right)^{-l}} \right],
\end{equation}
\end{figure*}
\begin{figure*}
\begin{equation}\label{eq:12}
       \tilde{U}_{P_k(e)}\cong\frac{1}{\log_2{M_k}}\sum_{m=1}^{\frac{1}{2}\prod_{i=1}^{K} M_{i}}  \left[ 1-\sum\nolimits_{l=0}^{L-1}{\binom{2l}{l}} {\left({1+2/\Upsilon_{1m}^{D_k}}\right)}^{-\frac{1}{2}} {\left(2\Upsilon_{1m}^{D_k}+4\right)^{-l}} \right]  , \quad M_k > 4. \qquad{}  
\end{equation}
\hrulefill
\end{figure*} 

\section{Numerical Results}
\begin{figure*}[htbp]
\centering

\subfloat[{}]{\includegraphics[width=0.66\columnwidth]{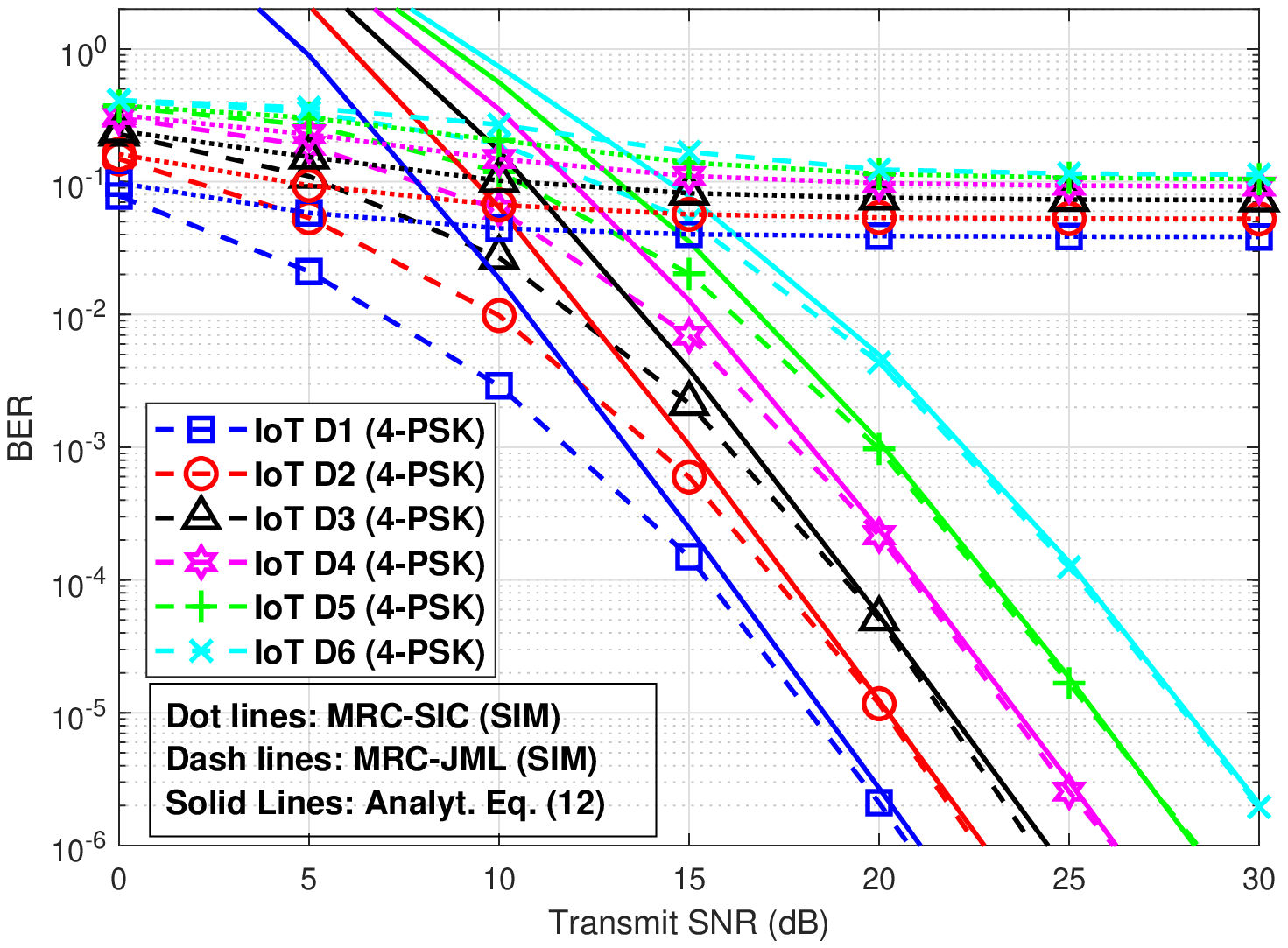}
\label{4D_4}}
\subfloat[]{\includegraphics[width=0.66\columnwidth]{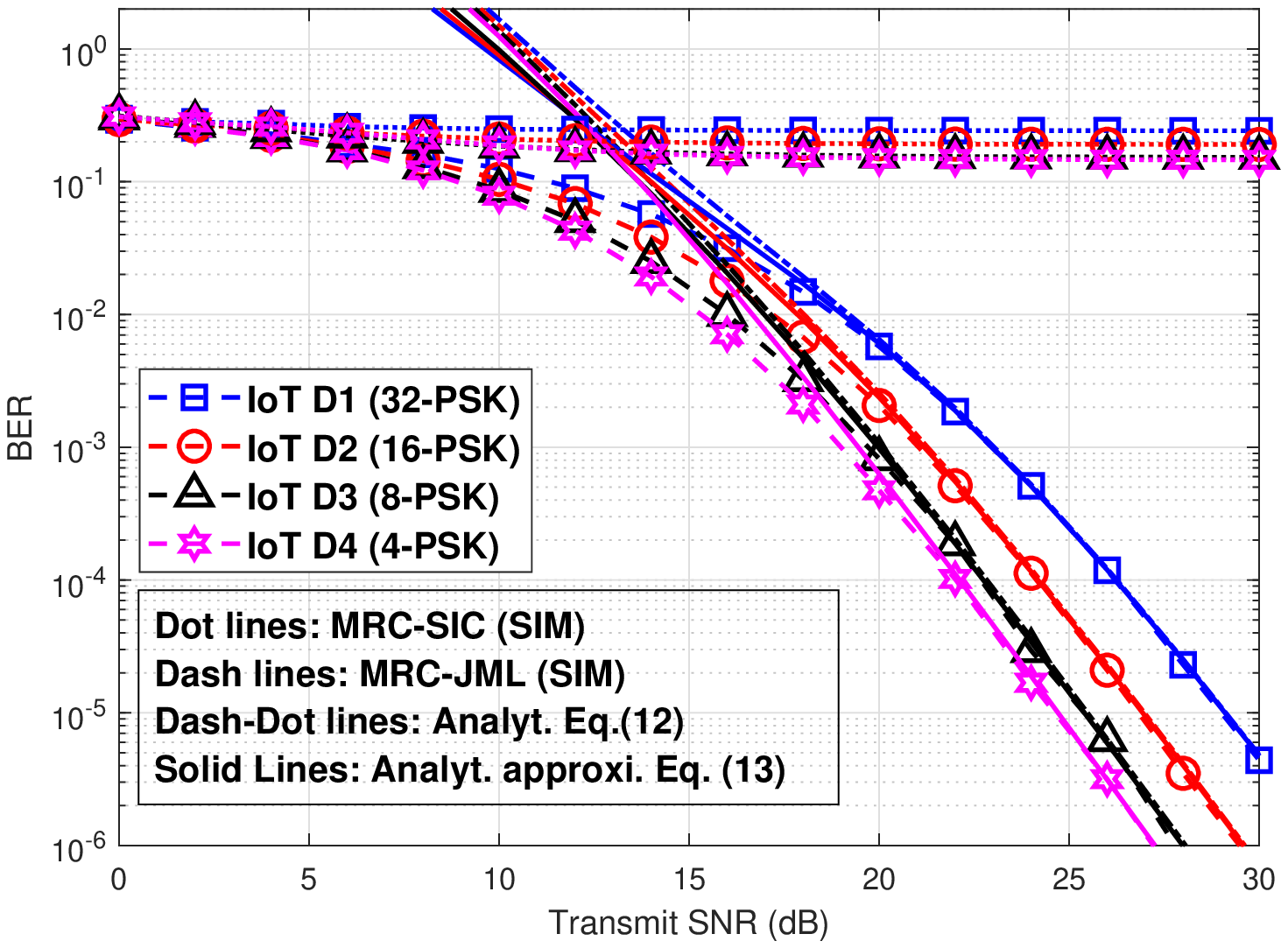}
\label{4D_ad}}
\subfloat[{}]{\includegraphics[width=0.66\columnwidth]{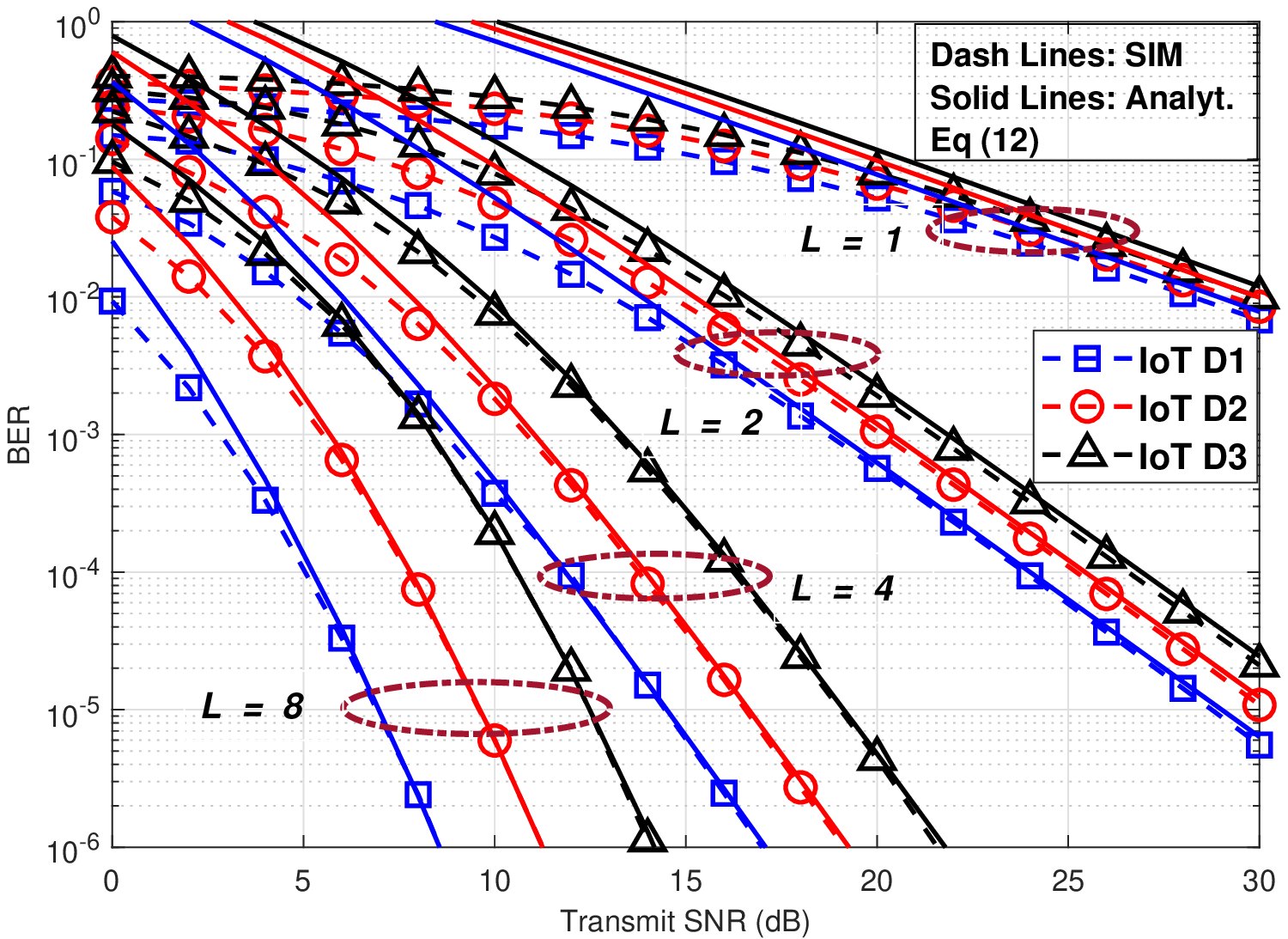}
\label{6D_ad}}
\caption{{Error performance of uplink IoT NOMA: BER vs Transmit SNR (dB). a)  $L=4$, $K=6$, $M_k=4, \ \forall k$. b) $L=4$, $K=4$, $M_1$ = $32$, $M_2$ = $16$, $M_3$ = $8$, $M_4$ = $4$  c) $L=1,2,4,8$, $K=3$, $M_k$ = $4, \ \forall k $.}}
\label{4D}
\end{figure*}

In this section, we validate the derived expressions by using computer simulations \footnote{We assume that $\epsilon_k=1,\ \forall k$, and $P_1=\cdots=P_K$, which is the worst case in terms of
interference. The channel conditions are varied according to the dominant gain channel by fixing variances as $\sigma_{k}^{2}$ = $2\sigma_{k+1}^{2}$ with $\sigma_{1}^{2} = 0$ dB.}. 
In Fig. \ref{4D_4} and Fig. \ref{4D_ad}, we present BER comparisons for MRC-JML and MRC-SIC with respect to SNR for $L=4$.
In Fig. \ref{4D_4}, we consider a fixed modulation where six ($K=6$) IoT devices emitting their data using $4$-PSK (i.e., $M_k=4,\ \forall k$) constellation. In Fig. \ref{4D_ad}, each device sends data  by using its own constellation order in a system composed of four ($K=4$) IoT devices. By exploiting the CSI provided by the BS via the broadcast signaling, commonly, to provide a considerable spectrum efficiency in adaptive modulation, a low order (stronger modulation) is chosen for the worst channel conditions, while high order (weaker modulation) is adopted for a high channel condition. Hence, the orders are chosen as follows: $M_1$=$32$, $M_2$=$16$, $M_3$=$8$, and $M_4$=$4$. In both scenarios, we can easily observe that the MRC-JML detector outperforms the MRC-SIC detector significantly. It is evident that the MRC-JML can remove completely the error floor in the high SNR regime. Differently, the MRC-SIC detector suffers greatly from the error floor and presents a poorer performance. Indeed, none of the symbols can be detected (i.e., $P_k(e)\sim0.5$) by using MRC-SIC. This clearly shows that the JML enables the MMTC by supporting higher number of devices (i.e., $K=4,6$) in the same resource block. Besides, it also allows using higher modulation orders, which is quite promising considering that the SIC detector fails to detect BPSK symbols even in two-user scenario \cite{Kara2018d,Kara2020}. Furthermore, the analytical results derived from expressions (\ref{eq:11})-(\ref{eq:12}) match well with the simulation results regardless of the number of users ($K$) or modulation orders ($M_k$) which validates our analysis. The upper bounds are very tight in the high SNR regime and it is also frequently for all upper bounds in communication systems. On the other hand, the upper bound is not tight in low SNR, and this is due to the fact that in this regime, the upper bound is a sum of PEPs with very large and close values. Moreover, once the performances of MRC-JML are compared according to the receiving SNR$_k$, the performances of devices with stronger gain channel surpass those of severe channels in the scenario of fixed modulation. Since the error patterns are the same (i.e., the same value of distances $\mathbf{E}_{kn}$ and $\mathfrak{d}_{k}$ which contribute to compute the PEP for each device $k$), this superiority is explained by the quality of signal-to-interference-plus-noise ratio SINR$_k$ for each device $k$. Whereas, when decreasing the modulation order for devices with a bad channel, it is noteworthy that the improvement of their performance is due to the compensation for the effect of the worst channel. This is caused by the enlarging of the minimum distance between symbols in the low order constellation.

Finally, to emphasis the benefit of diversity order provided by the JML detector, in Fig. \ref{6D_ad}, we present the BER performance of combining multiple signals at the BS for three ($K=3$) active IoT devices with different antenna ($L=1,2,4,8$) situations. As illustrated in Fig. \ref{6D_ad}, the performance improves each time the number of receiving antennas is increased owing to the improvement of quality of the link. We can also see that the theoretical results correspond well with the results of the simulation, this matching improves even at low SNR because of the enhancement in the power of the link due to the technique of combining at BS. Likewise, one can easily see that the MRC-JML reaches the full diversity order which corresponds to the number of antennas $L$. This order is noticed by checking the slope of BER curves in all figures. 
\section{Conclusion}
This paper presents a reliable multi-user uplink IoT scheme through SIMO-NOMA. By applying MRC-JML (i.e., optimum detector) instead of MRC-SIC (i.e., iterative detector commonly used in the literature) to detect data from several IoT devices, the proposed scheme has proven its effectiveness to discard the error floor. In addition, the JML attains the full diversity order and this confirms the superiority of the JML comparatively to SIC. A comprehensive analysis of BER based on union bound approach is studied. A tight upper bound of BER for multi-user is derived by assuming adaptive $M$-PSK modulation and arbitrary $K$ active IoT devices.

\bibliographystyle{IEEEtran}
\bibliography{Semira_WCL2021-1253_bib}
\end{document}